\documentclass[floatfix,twocolumn,showpacs,superscriptaddress,amssymb,array,aps,prb]{revtex4}

\usepackage{graphicx}

\newcommand{\vu}{Faculty of Sciences, Department of Physics and Astronomy, Condensed Matter Physics, Vrije~Universiteit,
De~Boelelaan~1081, 1081~HV~Amsterdam, The~Netherlands}
\newcommand{\lu}{Laboratory of Applied Optics, Department of Physics and Measurement Technology, Link\"oping University,
SE-581 83 Link\"oping, Sweden}

\begin{document}

\title{Optical properties of {MgH}$_2$ measured {\it in situ} in a novel gas cell for
ellipsometry/spectrophotometry}

\author{J. Isidorsson}
 \altaffiliation[Present address: ]{Solid State Physics, Uppsala University, Box 534, SE-751
21 Uppsala, Sweden} \affiliation{\vu}

\author{I.A.M.E.~Giebels}
 \email[Corresponding author. Electronic address: ]{giebels@nat.vu.nl}
 \affiliation{\vu}

\author{H. Arwin}
 \affiliation{\lu}

\author{R.~Griessen} \affiliation{\vu}

\date{\today}

\begin{abstract}
The dielectric properties of $\alpha$-MgH$_2$ are investigated in the photon energy range between 1 and 6.5~eV. For
this purpose, a novel sample configuration and experimental setup are developed that allow both optical transmission
and ellipsometric measurements of a transparent thin film in equilibrium with hydrogen. We show that
$\alpha$-MgH$_2$ is a transparent, colour neutral insulator with a band gap of $5.6 \pm 0.1$~eV. It has an intrinsic
transparency of about 80\% over the whole visible spectrum. The dielectric function found in this work confirms very
recent band structure calculations using the GW approximation by Alford and Chou.\cite{alford03} As Pd is used as
a cap layer we report also the optical properties of PdH$_x$ thin films.
\end{abstract}


\maketitle

\section{Introduction}
\label{sec:intro}

Among metal-hydrides\cite{kohlmann02} the magnesium hydrogen system has always occupied a special place.
Magnesium reacts reversibly with hydrogen to form MgH$_2$. It is thus considered to be one of the most
important candidates for the reversible storage of hydrogen\cite{schlapbach01b} due to its lightweight, low
cost and high hydrogen storage capacity (7.6 wt.\% of hydrogen). In spite of the large number of
publications on Mg-MgH$_2$ only little is known about the intrinsic physical properties of this system.
The scarcity of data for MgH$_x$ is mainly due to the experimental
difficulties encountered when trying to hydride Mg.\cite{selvam86} Nowadays, a great effort is made to
improve the hydrogen ab/desorption kinetics by making nanocrystalline Mg\cite{zaluska99} and/or adding e.g.
transition metals\cite{liang99,dehouche00,pelletier01} by ball milling.

Recent theoretical calculations\cite{vajeeston02} reproduce that MgH$_2$ undergoes various phase
transitions \cite{bastide80,bortz99} as a function of pressure. All theoretical calculations published so
far\cite{vajeeston02,yu88,pfrommer94,haussermann02} (using either the local density approximation (LDA) or
the generalized gradient approximation (GGA)) predict band gaps between 3.1 and 4.2~eV for $\alpha$-MgH$_2$.
This is smaller than the few sporadic experimental values reported until now. Krasko\cite{krasko82} mentions a value
of 5.16~eV for the band gap from unpublished work by Genossar. He and Pong\cite{he90} determined in an
indirect way using Penn's formula\cite{penn62} an average band gap of 5.8~eV. Yamamoto {\it et
al.}\cite{yamamoto02} report an optical transmission spectrum for a thin film in which the transmission
vanishes at 6.05~eV (205~nm). Apart from that Ellinger {\it et al.}\cite{ellinger55} found an index of
refraction of 1.95 and 1.96 for the ordinary and extraordinary rays at 589.3~nm. The dielectric properties
have not been studied at all. This triggered our interest to study the optical properties of MgH$_2$ in
detail.

Another strong reason for our interest in MgH$_2$ stems from metal-hydride switchable mirrors. In 1996
Hui-berts {\it et al.}\cite{huiberts96a} discovered that Y and La thin films change reversibly from
shiny, metallic films to transparent, insulating films upon hydrogenation either by changing the
surrounding hydrogen gas pressure or the electrolytic cell potential.\cite{notten96,kooij99} In 1997
Van~der~Sluis {\it et al.}\cite{sluis97} discovered that all rare-earth (RE) metals exhibit such a
switchable behaviour. However, all these materials have a characteristic colour in the fully
hydrogenated state because their band gap is in the visible ($E_g < 3$~eV). They showed that alloying
with Mg results in colour neutral switchable mirrors. This is very important for applications in e.g.\
`smart' windows. In 2001 Richardson {\it et al.}\cite{richardson01c} reported that Mg$_z$Ni ($z > 2$)
also features reversible switching behaviour upon hydrogenation. In all these cases (RE-Mg and
Mg$_z$Ni), the band gap is shifted to higher energies with increasing magnesium
concentration.\cite{sluis97,richardson01c,isidorsson01a,molen01,isidorsson01b} All these alloys
disproportionate upon hydrogenation.\cite{isidorsson01a,nagengast99b,divece02,divece03} This
disproportionation is also known for bulk RE-Mg.\cite{darriet80,sun02} The available data suggest that
the shift of the band gap to higher energies is due to the formation of MgH$_2$ which is expected to
have a large band gap. At the same time the reflectance in the low hydrogen phase (when the sample is
unloaded) increases due to Mg which has a high reflection.\cite{sluis97,giebels02a} At intermediate
concentrations the coexistence of Mg and MgH$_2$ seems to play an important role in the realization of a
highly absorbing, black state that is observed during loading and unloading of RE-Mg
alloys.\cite{griessen97b,giebels03a} It may also play a role in the black state observed in
Mg$_z$NiH$_x$.\cite{isidorsson02} Thus, to understand the role of MgH$_2$ in Mg-containing switchable
mirrors it is essential to determine the optical properties of MgH$_2$ thin films.

In this paper we study thin films of magnesium hydride with spectrophotometry and ellipsometry and determine
the dielectric function and the optical band gap. For this purpose we use a special substrate geometry and a novel
type of optical gas loading cell.

\section{Experimental}
\label{sec:exp}

MgH$_2$ thin films are made in two steps. First Pd capped metallic Mg films are deposited under UHV
conditions on an appropriate substrate. The Pd cap layer is necessary to protect Mg against oxidation and to
promote hydrogen dissociation and absorption. The films are subsequently loaded with hydrogen under high
pressure up to the composition MgH$_2$.

The hydrogenation of Mg to MgH$_2$ is, however, not straightforward as was shown by Krozer, Kasemo and
others.\cite{krozer87,krozer89,ryden89,krozer90,spatz93} Palladium capped Mg films exhibit unusual
kinetics due to the formation of a blocking MgH$_2$ layer at the interface between Pd and Mg. The
MgH$_2$ layer prevents H to diffuse\cite{luz80} to the metallic Mg that is still present underneath. The
formation of this blocking layer can be circumvented by starting hydrogenation at relatively low (1~mbar)
H$_2$ pressure at a temperature of 100$^{\circ}$C. Magnesium films with thicknesses up to 150~nm can be
fully transformed to MgH$_2$ in this way.\cite{ryden89,westerwaal02}

\subsection{Film deposition}
\label{subsec:deposition}

Thin, polycrystalline films of Mg and Pd are deposited at room temperature in an UHV MBE system with a
background pressure of 10$^{-9}$ mbar, using material of typically 99.9~\% purity. The magnesium films
are evaporated from a Knudsen cell and covered with a 10~nm thick Pd cap layer. These palladium films are deposited from
an e-beam evaporation unit. Typically, we deposit films simultaneously on
a 10x10~mm$^2$ glassy carbon substrate for Rutherford backscattering spectrometry (RBS), 10x10~mm$^2$
quartz substrates for X-ray diffraction (XRD), resistivity and/or AFM measurements, and on a quartz
substrate (\O~42~mm, Heraeus Suprasil 1) for optical measurements.

\subsection{Film characterization}
\label{subsec:characterization}

RBS is used to determine eventual contamination of the films. For this glassy carbon substrates are used in
order to separate the Mg signal from the background signal of the substrate. An oxygen contamination between $0.03
\leq [\mbox{O}]/[\mbox{Mg}] \leq 0.085$ has been found.

The thickness of the film is measured with a DekTac$^3$ or a Tencor Alpha step 200 mechanical stylus
profilometer. The surface structure, both before and after hydrogen loading, is investigated with a
NanoScope III atomic force microscope (AFM), operating in tapping mode using silicon cantilevers. The
scanned areas are typically 1x1 and 5x5~$\mu$m$^2$ from which the root-mean-square (RMS) roughness is
determined. The thickness and roughness values from these techniques are used as input parameters in the
modeling of the transmission, reflection and ellipsometric data (see Sec.~\ref{sec:results}).

Some samples are contacted ultrasonically with four 30~$\mu$m aluminium wires to monitor the resistivity with
the Van~der~Pauw method\cite{pauw58} during loading with hydrogen.

X-ray experiments are carried out with Cu K$\alpha$ radiation in a Rigaku `Rotaflex' RU 200 or Bruker
D8 Discover X-ray diffractometer to monitor the transformation of hcp Mg to rutile MgH$_2$ in a $\theta$-2$\theta$ mode.

\subsection{Optical techniques}
\label{subsec:optics}

Optical reflection and transmission measurements at room temperature (RT) are carried out in a Perkin Elmer
Lambda 900 spectrophotometer in the range $0.5 < \hbar\omega < 6.7$~eV ($2500 > \lambda > 185$~nm). The
specular and total transmission is recorded while the spectrophotometer is purged with argon or nitrogen in
order to reduce absorption by O$_2$ in the ultraviolet (UV), and H$_2$O in the infrared (IR). The quartz
substrates (without film) and Pd samples are measured in reflection geometry from the top side (i.e.\ the
metallic side) at near normal incidence (8$^{\circ}$) in an absolute reflection unit (using a so-called VN
geometry).

Ellipsometry measurements (at RT) in the energy range 1.0 to 6.5~eV ($1240 > \lambda > 190$~nm) are carried
out in a rotating analyzer variable-angle spectroscopic ellipsometer (VASE, J.A. Woollam Co. Inc.), using the
WVASE32 software program for data acquisition and analysis. This instrument measures the ratio of the complex
Fresnel reflection coefficients $R$ of parallel ($p$) and perpendicular ($s$) polarized light.\cite{azzam77}
This ratio defines the ellipsometric angles $\Psi(\omega)$ and $\Delta(\omega)$ according to
\begin{equation}
\label{eq:fresnel} \frac{R_p}{R_s} = \tan(\Psi(\omega))\exp(i\Delta(\omega))
\end{equation}

Three angles of incidence (60, 65 and 70$^{\circ}$) are used to obtain adequate sensitivity over the whole
spectral range.  Standard deviation and ellipsometric data ($\Psi,\Delta$) are recorded at each data point as
an average of at least 100 revolutions, and up to 4000 revolutions of the analyzer for the most critical data.

\subsection{Semi-cylindrical substrate}
\label{subsec:substrate}

As the Pd cap layer on top of the very transparent MgH$_2$ layer is strongly absorbing, ellipsometry cannot
be carried out from the Pd-side. Thus, ellipsometry measurements need to be performed from the `backside',
through the substrate. Flat substrates would need to be so thick that reflections from the front and
backside of the substrate can be well separated. With a 2~mm diameter of the light beam the substrate must
be thicker than 3 mm. However, at large angles of incidence the intensity loss in the light beam is
substantial due to reflections at the air/substrate interface. Furthermore, at energies close to the limit
of the ellipsometer (6-6.5~eV), the intensity of the light beam is diminishing quickly. These limitations
can be avoided with a semi-cylindrical substrate.

\begin{figure}
\includegraphics[height=9cm]{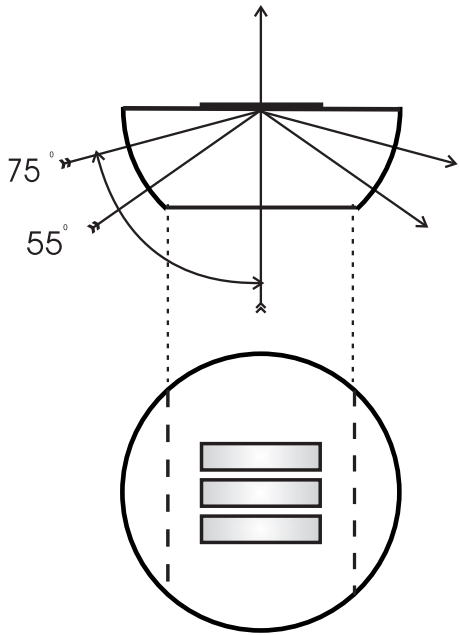}
\caption{Semi-cylindrical quartz glass substrate for ellipsometry and transmission measurements. The upper
figure is a side view of the substrate, the one on the bottom a top view. A perspective view is given in
Fig.~\ref{fig:novelcell}. The angle of incidence of light can be
varied between 55 and 75$^{\circ}$. On the flat part of the substrate three samples with different
Mg thicknesses are deposited. They are all covered with a 12~nm thick Pd cap layer.} \label{fig:substrate}
\end{figure}

This substrate is designed for normal incidence of light at the ambient/substrate
interface and oblique incidence at the internal substrate/MgH$_2$ interface (see Fig.~\ref{fig:substrate}).
For the normal incident approximation to be valid, the diameter of the semi-cylindrical substrate must be large
compared to the size of the light beam (2~mm). For practical reasons we choose a semi-cylindrical substrate
with a diameter of 42~mm. The top part of the semi-cylindrical substrate is cut away parallel to the base
surface to enable transmission measurements. This design allows the angle of the incident light onto the sample
to vary between 55$^{\circ}$ and 75$^{\circ}$ from the normal. Both the flat and the
semi-cylindrical substrates are made of quartz glass (Heraeus Suprasil 1). This material is transparent deep
into the UV beyond the limit of our ellipsometer and spectrophotometer.

On the top of the large flat face we deposit, under exactly the same condition, three films with
different Mg thicknesses (see Fig.~\ref{fig:substrate}). This allows us to analyze compositionally identical
films. This method makes the determination of the dielectric function from ellipsometric data more
reliable.\cite{mcgahan93,jarrendahl98}

\subsection{Optical gas loading cell and high pressure loading chamber}
\label{subsec:novelcell}

In order to measure the optical properties of MgH$_2$ and PdH$_x$ {\it in situ} in equilibrium with hydrogen
at various pressures, we designed an experimental setup consisting of three parts: i) a substrate/window (see
Fig.~\ref{fig:substrate}), ii) an optical gas loading cell, and iii) a high pressure loading chamber.
Components ii) and iii) are described below in more detail. The setup also includes a special substrate
holder with a sliding mask for the deposition of the films, and a holder to attach the optical gas loading
cell to the ellipsometer.

\begin{figure}
\includegraphics[width=9cm]{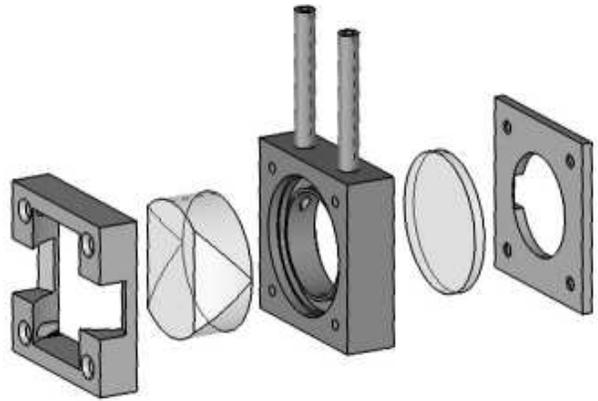}
\caption{Exploded view of the optical gas loading cell for {\it in situ} ellipsometry and transmission
measurements in a hydrogen environment. In this sketch the cell is shown with the semi-cylindrical
substrate (see Fig.~\ref{fig:substrate}) as sample window at the front side. Alternatively, a flat window
such as the one at the backside of the cell, can be used as sample substrate. \label{fig:novelcell}}
\end{figure}

A semi-cylindrical or a flat substrate with the Pd-capped MgH$_2$ films deposited on top functions as window
in the optical gas loading cell (see Fig.~\ref{fig:novelcell}).
For ellipsometry the sample/window is
illuminated through the substrate to measure the `backside' of the film. For {\it in situ} transmission
measurements a window is mounted on the opposite side of the cell as well. The sample can be exposed to a
controlled hydrogen gas atmosphere during the measurements via two tubes connected to a vacuum pump and
hydrogen gas cylinder. The cell is designed for vacuum, but works also reliably up to a few bar hydrogen
pressure.

\begin{figure}
\includegraphics[height=10cm]{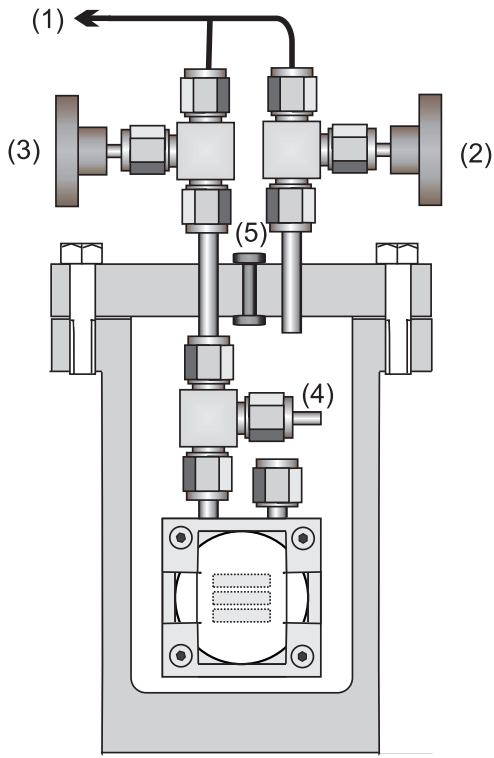}
\caption{High pressure chamber to load the Mg samples to MgH$_2$ at a pressure of 100~bar H$_2$. The
optical gas loading cell shown in Fig.~\ref{fig:novelcell} is mounted inside the chamber. The various parts
are: (1) connection to the vacuum and gas handling system, (2) a valve to open and close the high pressure
chamber, (3) and (4) valves to open and close the optical gas cell, (5) is the electrical feed-through to
enable {\it in situ} measurements of the resisitivity and temperature. \label{fig:highpres}}
\end{figure}

As already mentioned in the introduction of Sec.~\ref{sec:exp}, hydrogenation of Mg to MgH$_2$ can be
successfully achieved at moderate temperatures ($100^{\circ}$C) by starting at low hydrogen
pressure.\cite{ryden89,westerwaal02} Therefore, to be sure that our thin films of Mg are completely
transformed
to MgH$_2$ we start loading with a H$_2$ pressure of 1~mbar and increase it in steps (within a few
hours) up to 100~bar H$_2$. To do this the optical gas loading cell is mounted inside a high pressure
loading chamber (see Fig.~\ref{fig:highpres}).
This chamber is made of stainless steel and
proof-pressurized to 200~bar. The design is such that the pressure in the high pressure chamber is
everywhere the same, and it is not necessary to expose the sample to the ambient after loading at 100~bar
H$_2$. It is possible to release hydrogen and close the optical gas loading cell with valve (3) when
a pressure of 1~bar hydrogen or less is reached. With valve (2) closed as well, the system is disconnected
from the gas tubes (1) (while still filled with H$_2$). Afterwards, valve (2) is opened to equalize the
pressure outside and inside the high pressure chamber. Then, the chamber is opened, and the valve on top of
the optical gas cell (number (4)) is closed. The cell can then be disconnected from the lid of the high
pressure loading chamber while still filled with hydrogen. The high pressure loading chamber is also
equipped with an electrical feed-through (5) with several wires to permit measurements of the resistivity
of a sample during hydrogenation and of the temperature inside the chamber with a RhFe100 sensor. During
hydrogen loading of a Mg film the total chamber can be resistively heated up to 100$^{\circ}$C.

\section{Results}
\label{sec:results}

\subsection{Sample characterization}
\label{subsec:charac}

Since Mg is a metal and MgH$_2$ an insulator, the time evolution of hydrogenation can be followed {\it in
situ} in real time by monitoring the change of the resisitivity.\cite{hjort96a} This allows us to optimize
the hydrogen pressure (starting at low pressures and increasing it stepwise to 100~bar H$_2$) in such a way
that no impenetrable MgH$_2$ layer is formed at the interface between Pd and Mg. For practical reasons we
have mounted an extra sample in the high pressure loading chamber for resistivity measurements.

The resistivity of this as-deposited 150~nm Mg film covered with 15~nm of Pd at RT is 6.5~$\mu\Omega$cm
(the literature value for bulk Mg at 20$^{\circ}$C is 4.4~$\mu\Omega$cm\cite{handbook01}). The reflection
of this as-deposited film is high in the visible and near-infrared regions ($\sim$80\%). Both the low
resistivity and the high reflection indicate the good quality of the film. After loading the resistivity
reaches 680~$\mu\Omega$cm under 100~bar H$_2$ at 100$^{\circ}$C. Since MgH$_2$ is an insulator one would at
first sight expect a much higher value. The moderate resistivity found experimentally is, however, due to
the metallic Pd cap layer that shortcuts the MgH$_2$ layer. Moreover, at a temperature of 100$^{\circ}$C Mg and Pd
may interdiffuse to form a Mg-Pd alloy.\cite{krozer90,fischer91} This intermixing has been suggested for Pd
capped Y as well,\cite{molen99} and was conclusively shown with photoelectron spectroscopy
recently.\cite{borgschulte01} RBS showed an intermixing of Mg and Pd in our films as well. This can be due
to either alloying, interface roughening or both. The net result is that a relatively Pd-rich Pd-Mg alloy
is formed on top of MgH$_2$ that absorbs some hydrogen but does not become insulating and this causes the
shortcut.

\begin{figure}
\includegraphics[width=8cm]{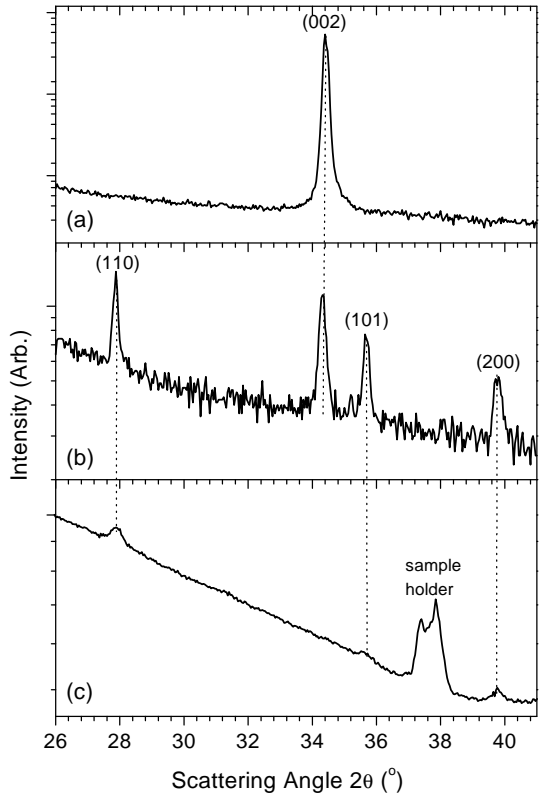}
\caption{X-ray diffraction spectra of a 188~nm Mg/10~nm Pd film (a) as-deposited, (b) loaded up to 1~bar
H$_2$ and (c) of a 150~nm Mg/15~nm Pd film loaded up to 100~bar H$_2$. The large background is due to the
quartz substrate. \label{fig:xrd}}
\end{figure}

In the as-deposited metallic state, hcp Mg has a preferential growth direction, and only the (002)
reflection is present in the X-ray diffractogram (see Fig.~\ref{fig:xrd}(a)). Loading a thin Mg film in
1~bar H$_2$ at 100$^{\circ}$C does not transform Mg completely to MgH$_2$ (see Fig.~\ref{fig:xrd}(b)).
Loading at 100~bar and 100$^{\circ}$C, on the other hand, left no traces of metallic Mg. Only the peaks
corresponding to the tetragonal structure of the rutile type\cite{ellinger55} of $\alpha$-MgH$_2$ are
observed (see Fig.~\ref{fig:xrd}(c)). Such a preferred growth direction is not observed for
MgH$_2$ where weak signals from the (110), (101) and (200) peaks can be seen. Rocking curves around the
(002) Mg peak and the (110) MgH$_2$ peaks show that our samples are polycrystalline.

\begin{figure}
\includegraphics[height=10cm]{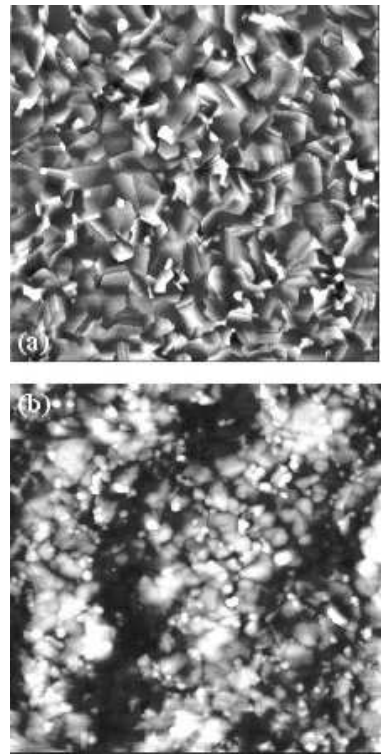}
\caption{AFM images of the surface (5x5~$\mu$m$^2$) of a 101~nm Mg/12~nm Pd (a) as-deposited (b) after
loading with hydrogen at 100~bar. The root mean square roughness increases from 5 to 14~nm due to the 32\%
volume expansion accompanied by the transition from hcp Mg to rutile MgH$_2$. \label{fig:AFM}}
\end{figure}

AFM measurements revealed a significant difference between the as-deposited Pd covered Mg film and the
fully hydrogenated films (see Fig.~\ref{fig:AFM}).
Mg expands by 32\% in volume when transforming from hcp Mg to
rutile MgH$_2$.\cite{schober81} Since the film is clamped by the substrate it cannot expand laterally and
all the expansion must take place out-of-plane. With AFM we indeed noted an increase in the RMS roughness
from 5 nm to 14 nm. It can be seen as well that our top layer of Pd is cracked. With a mechanical stylus
profilometer we found a corresponding increase of the thickness of the film from 113 to 162~nm. This 43\%
increase is larger than the expected 32\% volume expansion because the mechanical stylus has a tip radius
of 12.5~$\mu$m, and hence cannot probe the deep valleys seen on the AFM image. During ellipsometry we look
through the substrate at the backside of our films and not from the top side as with AFM. Nevertheless, we found that
surface roughening of our Pd top layer needed to be taken into account when modeling the ellipsometric
data.

\subsection{Transmission and band gap of MgH$_2$}
\label{subsec:trans}

\begin{figure}
\includegraphics[width=8cm]{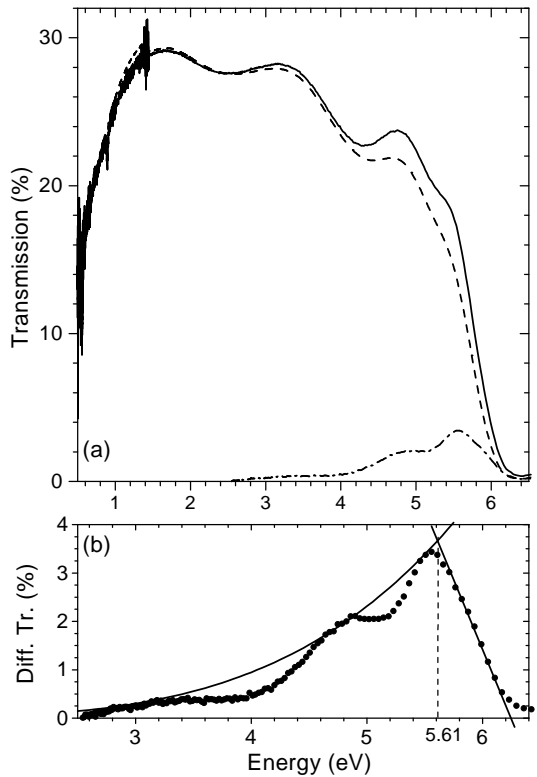}
\caption{(a) Total (solid line), specular (dashed line) and diffuse (dashed dotted line) transmission as a
function of photon energy for a 150~nm thick MgH$_2$ film capped with 12~nm Pd and loaded at 100$^{\circ}$C
in 100 bar of hydrogen. (b) Detail of the diffuse transmission. A fit using Eq.~\ref{eq:rayleigh} and an
extrapolation of the transmission edge are shown. The intersection of these two curves gives an estimate of
5.61~eV for the optical band gap. \label{fig:transMgH2}}
\end{figure}

The optical transmission of Pd capped MgH$_2$ films is measured {\it in situ} in 1~bar H$_2$ using the gas
loading cell (see Fig.~\ref{fig:novelcell}). Figure~\ref{fig:transMgH2} shows the total, specular and
diffuse transmission of a 150~nm thick MgH$_2$ film capped with 12~nm Pd, loaded at 100~bar H$_2$ and
100$^{\circ}$C.
The total transmission is measured with the optical gas loading cell placed at the entrance port of the
integrating sphere in the spectrophotometer. In this experiment a flat 3~mm thick quartz glass substrate
is used. Since we look at our film from the substrate side, the MgH$_2$ layer is situated 3~mm away from
the port of the integrating sphere. The specular transmission is measured with the sample in the sample
compartment, using the direct detector to monitor the signal. The difference between these two signals
is the diffuse (scattered) transmission. It is probably due to the rough surface (see
Sec.~\ref{subsec:charac} and Fig.~\ref{fig:AFM}(b)) of our loaded samples. This diffuse transmission,
$T_d$, has a strong wavelength dependence and is proportional to
\begin{equation} \label{eq:rayleigh} T_d \propto
\frac{1}{\lambda^4} \propto (\hbar\omega)^4,
\end{equation}
where $\lambda$ is the wavelength of light.\cite{rayleigh71} It is clear from
Fig.~\ref{fig:transMgH2}(b) that the diffuse transmission decreases strongly above the band gap as the
film starts to absorb light. The optical band gap $E_{\mbox{\tiny{g}}}$, can be estimated from the
intersection of a fitted $(\hbar\omega)^4$ curve to the data and an extrapolation of the flank of the
absorption edge. Using this so called `Rayleigh method' we find $E_{\mbox{\tiny{g}}} = 5.61$~eV for this
sample and $E_{\mbox{\tiny{g}}} = 5.67$~eV for a second sample.

Another estimate for $E_g$ can be obtained from the absorption edge of the transmission spectra
using the Lambert-Beer law, $T(\omega) = T_0 \exp[-\alpha(\omega) d]$, where $\alpha$ is the absorption
coefficient, $d$ the film thickness and $T_0$ contains the transmission of the Pd cap layer and the quartz
substrate. In the region of the absorption edge $T_0$ can be considered as constant in our films (see
Figs.~\ref{fig:nkquartz} and \ref{fig:eps12PdHx}). The frequency dependence of $\alpha$ near the band edge
is related to the optical gap through,\cite{tauc66,johnson67}
\begin{equation} \label{eq:tauc}
\alpha(\omega) \propto \frac{(\hbar\omega - E_{\mbox\tiny{g}})^\nu}{\hbar\omega}.
\end{equation}
For direct, allowed (forbidden) transitions $\nu = \frac{1}{2}$ ($\nu = \frac{3}{2}$) and for indirect, allowed
(forbidden) transition $\nu = 2$ ($\nu = 3$). In amorphous material it has been found that $\nu = 2$ gives the best
results. Combining these equations gives
\begin{equation}
\label{eq:lnT} \ln T(\omega) = \ln T_0 - C\frac{(\hbar\omega - E_g)^{\nu}}{\hbar\omega},
\end{equation}
and the constants $\ln T_0$, $C$, and $E_g$ are determined from a fit to the spectra near the transmission
edge in the interference-free region.

Applied to the total transmission this 'Tauc procedure' gives a gap of $5.48 \pm 0.05$~eV using $\nu
=2$. It was also possible to get a Tauc fit with $\nu = 3$ and $\nu = \frac{3}{2}$. However, using $\nu
= 3$ gives values that are too low compared to the `Rayleigh procedure' and we might be fitting an
interference fringe instead of the absorption edge. For $\nu = \frac{3}{2}$ the quality of the fit is
not as good and we obtain a gap of $5.75 \pm 0.05$~eV from the total transmission.

Since the diffuse transmission increases rapidly at small wavelengths (i.e.\ with increasing energy), it
is clearer where the absorption starts in this spectrum than in the total transmission. Therefore, we
conclude that the band gap of MgH$_2$ is $5.6 \pm 0.1$~eV.

\subsection{Ellipsometry}
\label{subsec:ellipsometry}

\subsubsection{Modeling strategy}
\label{sssec:modelstrat}

Extracting the dielectric function of a layer from ellipsometric data on samples like ours, which consists
of several thin layers on a substrate, is a complex task. The complicated inversion of  the ellipsometric
data to the dielectric function is, however, greatly simplified if the optical properties of each individual
layer is measured in separate experiments. For this reason we adopted the following strategy to determine the
dielectric function of MgH$_2$.

We start by investigating the dielectric properties of the quartz substrate. The second step is to evaluate
the optical properties of the hydrogenated Pd cap layer. This is done by investigating a 12 nm thick film of
Pd on quartz, hydrogenate it in 1~bar H$_2$ using the optical gas loading cell and measure it in the
ellipsometer. In a third step we study the properties of the hydrogenated interface region between Pd and
MgH$_2$ carefully since earlier experiments with Mg-Pd thin films showed that interdiffusion starts already at
100$^{\circ}$C.\cite{krozer90,fischer91} This is done by investigating a
`Pd-Mg' alloy layer consisting of 10 nm Mg  on a quartz substrate covered with 10 nm Pd. This sample is
loaded with hydrogen at 100$^{\circ}$C and 100~bar and measured in the ellipsometer. The final step is to
measure the total stack (quartz, MgH$_2$, Pd-Mg) and to extract the optical properties of MgH$_2$ using
the optical properties of all the other layers.

An optical model is defined for each sample and in the fitting procedure the difference between the
calculated (cal) and the measured (exp) ($\Psi,\Delta$)-values (see Sec.~\ref{subsec:optics}) are
weighted with the experimental standard deviation $\sigma$ and fitted with a Levenberg-Marquardt
algorithm to minimize the mean squared error (MSE) according to\cite{jellison91,press88}

\begin{widetext} \begin{eqnarray} \label{eq:mse} \mbox{MSE}  =
\frac{1}{2N-M}\sum_{\lambda}\sum_{\Theta}\left\{\left[\frac{\Psi_{\lambda,\Theta,\mbox{\tiny{cal}}}-\Psi_{\lambda,\Theta,\mbox{\tiny{exp}}}}{\sigma_{\lambda,\Theta,\Psi}}\right]^2
+
\left[\frac{\Delta_{\lambda,\Theta,\mbox{\tiny{cal}}}-\Delta_{\lambda,\Theta,\mbox{\tiny{exp}}}}{\sigma_{\lambda,\Theta,\Delta}}\right]^2\right\}
\end{eqnarray} \end{widetext} where $N$ is the number of ($\Psi,\Delta$)-pairs, $M$ is the number of
fitting parameters, and the indices $\lambda$ and $\Theta$ denote data points at different wavelengths and
angles. In most cases also normal incidence transmission data, $T$, are used to improve the accuracy of the
determination of the dielectric function.\cite{johs94} Then, a third term is included in the summation in
Eq.~\ref{eq:mse}.

In the modeling we
take into account experimental errors in incident angle and angular spread due to the
substrate design, and film thickness non-uniformity. It is difficult to model both the thickness and the dielectric
properties simultaneously in ellipsometry.\cite{mcgahan93,jarrendahl98} Thus, we allow the layer thicknesses
to vary only slightly around our measured thickness values during fitting. The output from the modeling
consists of the best-fit value of the Lorentz-Drude parameters (see Eq.~\ref{eq:LD}) and their 90\% confidence intervals.

\subsubsection{Optical constants of the glass substrate}
\label{sssec:optconstsubstr}

\begin{figure}
\includegraphics[width=8cm]{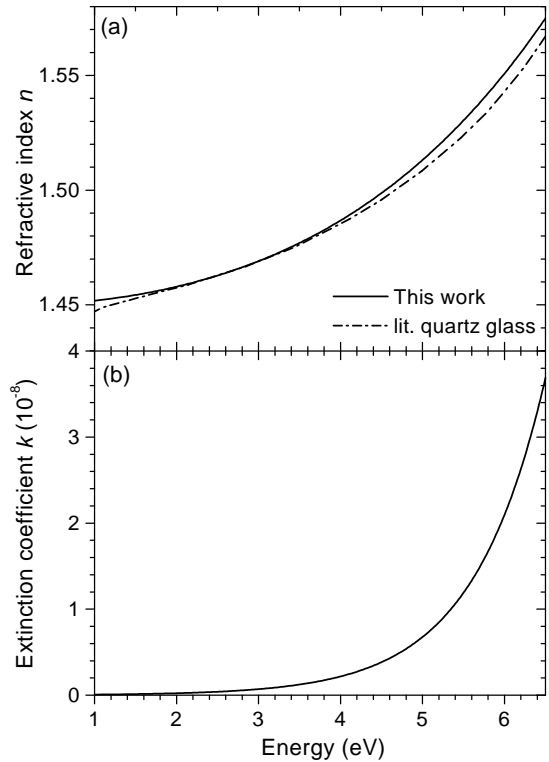} \vspace{0.2cm}
\caption{(a) Refractive index, $n$ and (b) extinction coefficient, $k$ (b) of the quartz substrates as
determined from ellipsometry and reflection and transmission measurements. For comparison literature
data for the refractive index of quartz glass in Ref.~\onlinecite{palik98} are shown as a dot-dashed
line. \label{fig:nkquartz}}
\end{figure}

The optical constants of the quartz substrates (both flat and semi-cylindrical) and the quartz window used
in the optical gas loading cell (see Fig.~\ref{fig:novelcell}) are determined using optical reflection and
transmission measurements and ellipsometry. This is straightforward, and our results match the tabulated
values of the refractive index from the manufacturer as well as those of quartz glass cited in
Ref.~\onlinecite{palik98} (see Fig.~\ref{fig:nkquartz}(a)). The extinction coefficients being not tabulated
in Ref.~\onlinecite{palik98} are assumed to be zero. Our results on the extinction coefficient show a
slight absorption near and above 6.5~eV, but still below  $10^{-7}$ (see Fig.~\ref{fig:nkquartz}(b)). The
corresponding dielectric function is used in the consecutive modeling of the metal hydride layers.

\subsubsection{Optical properties of PdH$_x$}
\label{sssec:optconstPdH}

\begin{figure}
\includegraphics[width=8cm]{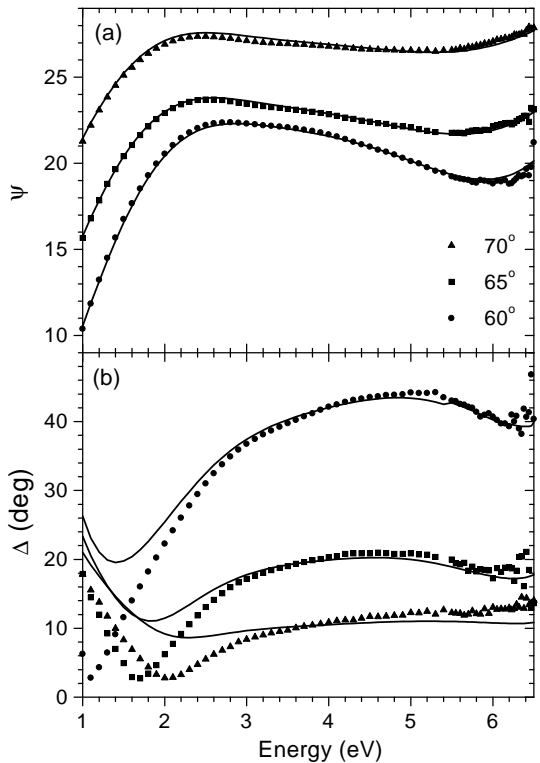}
\caption{Experimental and fitted ellipsometric data, $\Psi$, (a) and $\Delta$ (b) for a 12~nm thick Pd
film in 1 bar H$_2$ used to determine the dielectric function of PdH$_x$ (see Fig.~\ref{fig:eps12PdHx}).
The mean square error corresponding to the fit is 7.4. \label{fig:psideltaPdHx}}
\end{figure}

A 12~nm thick Pd film, deposited on quartz is exposed to 1~bar hydrogen at RT in the optical gas loading
cell and investigated in the ellipsometer (see Fig.~\ref{fig:psideltaPdHx} for the experimental and fitted
data).
The Pd hydride, PdH$_x$, that is formed is a strongly absorbing metal. Its dielectric function
$\epsilon(\omega) = \epsilon_1 + i \epsilon_2$ can be adequately parameterized with a Lorentz-Drude (LD)
model:
\begin{equation} \label{eq:LD} \epsilon(\omega)  = \epsilon_{\infty} - \sum_{i=1}^{N}
\frac{\omega_{p,i}^2}{\omega^2 + i \omega / \tau_i} + \sum_{j=1}^{M} \frac{f_j}{\omega_j^2 - \omega^2 - i
\Gamma_j \omega}
\end{equation}
where the constant $\epsilon_{\infty}$ accounts for excitations far above
6.5~eV; the $N$ Drude terms describe the free-carrier response with $\omega_{p,i}$ the plasma frequency of
the $i$-th Drude term and $\tau_i$ the relaxation time; the $M$ Lorentz terms represent the effect of
interband transitions with $f_j$ the intensity of the $j$-th oscillator, $\omega_j$ its energy, and
$\Gamma_j$ its broadening. The relation between the dielectric function and the refractive index, $n$ and
extinction coefficient, $k$, is: $\epsilon_1 = n^2 - k^2$ and $\epsilon_2 = 2nk$. The LD parameters obtained
for PdH$_x$ in 1~bar H$_2$ are given in Table~\ref{table:pdhx1bar}.
\begin{table*}

\caption{\label{table:pdhx1bar}Lorentz-Drude parameters and their 90\% confidence intervals of a 12~nm
thick PdH$_x$ layer in 1 bar H$_2$ obtained from ellipsometric data (see Fig.~\ref{fig:psideltaPdHx}).
MSE = 7.4, $\epsilon_{\infty}$ = 1.266 $\pm$ 0.371. All parameters are in eV.}

\begin{ruledtabular} \begin{tabular}{lllllll}

($i$) & $\omega_{p,i}$ & $1/\tau_i$ & ($j$) & $\omega_j$ & $\sqrt{f_j}$ & $\Gamma_j$  \\ \hline (1) &
3.389 $\pm$ 0.901 & 0.1892 $\pm$ 0.019 & &&& \\ (2) & 8.656 $\pm$ 2.02 & 1.775 $\pm$ 0.147 & &&& \\
  &&& (1) & 3.418 $\pm$ 0.124 & 8.588 $\pm$ 4.6 & 5.195 $\pm$ 0.691 \\
  &&& (2) & 6.878 $\pm$ 0.119 & 11.32 $\pm$ 4.63 & 7.713 $\pm$ 0.519 \\
  &&& (3) & 9.669 & 11.96 & 0.5601 \\

\end{tabular} \end{ruledtabular} \end{table*}

In addition to the sample for ellipsometry, an identical sample on a flat substrate is prepared for
measurements in the spectrophotometer. The transmission and absolute reflection are determined for this
sample in 40~mbar hydrogen (4\% H$_2$ in Ar). At higher pressures it is not possible to use the optical
gas loading cell when we determine the absolute reflection since we must look directly at the Pd sample
and not via the substrate. To obtain a hydrogenated Pd sample, the whole spectrophotometer is purged in
Ar containing 4\% H$_2$, corresponding to a partial pressure of 40~mbar H$_2$. This is the highest H$_2$
concentration we can use in the (open) spectrophotometer.
\begin{figure}
\includegraphics[width=8cm]{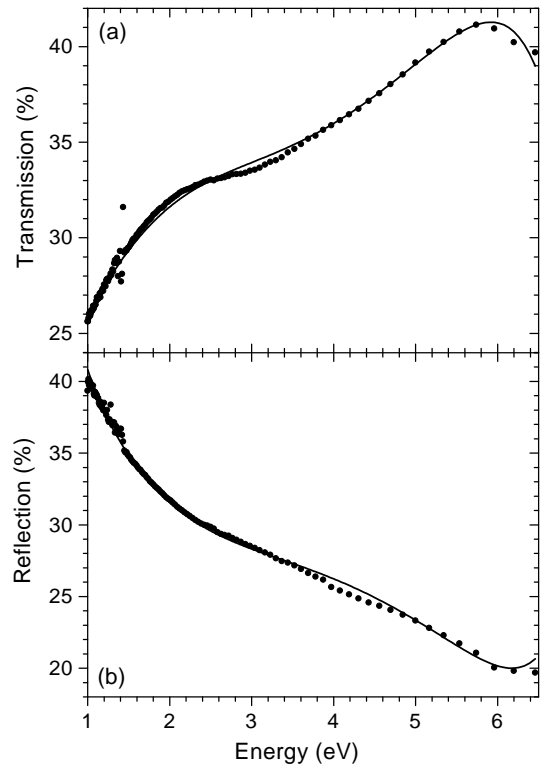}
\caption{Experimental and fitted transmission (a) and reflection (b) data for a 12~nm thick Pd film in
40 mbar H$_2$ (4\% H in Ar) used to determine the dielectric function of PdH$_x$ (see
Fig.~\ref{fig:eps12PdHx}). The mean square error corresponding to the fit is 1.04.
\label{fig:refltransPdHx}}
\end{figure}
In Fig.~\ref{fig:refltransPdHx} the experimental and fitted data are displayed, the LD parameters
obtained for PdH$_x$ in 40~mbar H$_2$ are given in Table~\ref{table:pdhx40mbar}.
\begin{table*}

\caption{\label{table:pdhx40mbar}Lorentz-Drude parameters and their 90\% confidence intervals of a 12~nm
thick PdH$_x$ layer in 40~mbar H$_2$ (4\% H$_2$ in Ar) obtained from reflection and transmission data
(see Fig.~\ref{fig:refltransPdHx}). MSE = 1.04, $\epsilon_{\infty}$ = 1.280 $\pm$ 0.241. All parameters
are in eV.}

\begin{ruledtabular} \begin{tabular}{lllllll}

($i$) & $\omega_{p,i}$ & $1/\tau_i$ & ($j$) & $\omega_j$ & $\sqrt{f_j}$ & $\Gamma_j$  \\ \hline (1) &
5.156 $\pm$ 0.955& 0.001 $\pm$ 0.0241 & &&& \\ (2) & 8.382 $\pm$ 2.65 & 2.238 $\pm$ 0.323 & &&& \\
  &&& (1) & 4.131 $\pm$ 0.111 & 9.815 $\pm$ 4.17 & 7.104 $\pm$ 0.581 \\
  &&& (2) & 7.623 $\pm$ 0.249 & 7.793 $\pm$ 4.14 & 0.927 $\pm$ 0.235 \\

\end{tabular} \end{ruledtabular} \end{table*}

The difference in the MSE (see Table~\ref{table:pdhx1bar} and \ref{table:pdhx40mbar}) between the 1~bar
and 40~mbar measurements is mainly due to the different substrate geometry, semi-cylindrical vs. flat
substrate. We assign the major part of the difference in the MSE to the cylindrical incident and exit
surface of the semi-cylindrical substrate and to the fact that the ellipsomety measurements have been
performed from the `backside' of the sample through the substrate. Note that this does not change the
dielectric function, it merely gives a larger spread in the input data for the analysis.

\begin{figure}
\includegraphics[width=8cm]{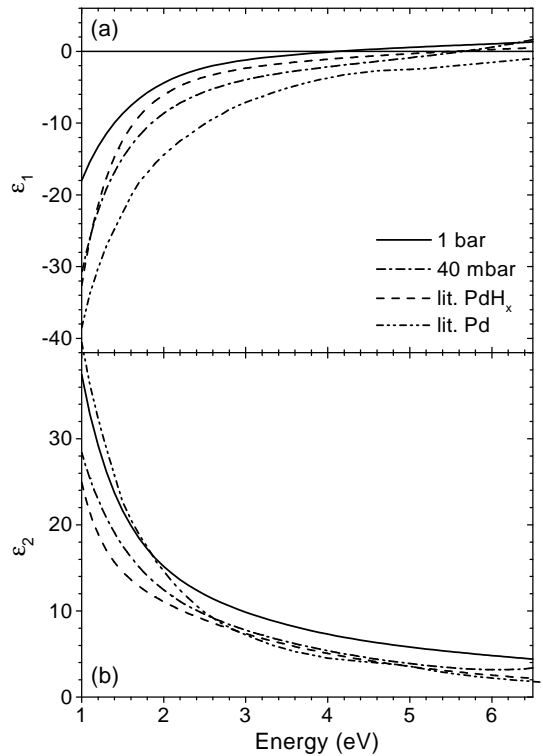}
\caption{Real (a) and imaginary (b) parts of $\epsilon(\omega)$ for the PdH$_x$ films as determined from
ellipsometry in 1 bar H$_2$ (see Fig.~\ref{fig:psideltaPdHx}) and reflection and transmission
measurements in 40~mbar H$_2$ (4\% H$_2$ in Ar) (see Fig.~\ref{fig:refltransPdHx}). For comparison, the
dielectric function found by Rottkay {\it et al.}\cite{rottkay99a} for PdH$_x$ in 4\% H$_2$ and the
dielectric function of pure Pd according to Ref.~\onlinecite{palik98} are included.
\label{fig:eps12PdHx}}
\end{figure}

In Fig.~\ref{fig:eps12PdHx} we show the resulting dielectric
function at 1 bar, and 40~mbar H$_2$ partial hydrogen pressure, and compare it to literature data for PdH$_x$
by Rottkay {\it et al.}\cite{rottkay99a} and literature data for Pd from Ref.~\onlinecite{palik98}.

The plasma frequency, $\omega_p$, of PdH$_x$ in 1~bar H$_2$ is slightly larger than the one in 40~mbar
H$_2$. Because $\omega_p \propto \sqrt{n_c}$ with $n_c$ the charge carrier density, there are more free
charge carriers in PdH$_x$ in 1~bar than in 40~mbar H$_2$. Values for the (optical) resistivity $\rho$
can be derived from the Drude parameters using
\begin{equation} \label{eq:optres}
\rho_{\mbox{\tiny{opt}}} = \frac{1}{\epsilon_0 \omega_p^2 \tau},
\end{equation}
with $\epsilon_0$ the vacuum permittivity, $\omega_p$ the plasma frequency and $\tau$ the electron relaxation
time. This optical resistivity calculated
for the dominant Drude term no.~2 shows the same trend as the plasma frequency itself: In 40~mbar H$_2$ PdH$_x$ has
a resistivity of 235~$\mu\Omega$cm, in 1~bar 178~$\mu\Omega$cm. However, bulk Pd has a resistivity of 10.53~$\mu\Omega$cm
and PdH$_x$ has a maximum resisitivity of about 20~$\mu\Omega$cm when $x=0.7$ at RT.\cite{geerken83} Our much
larger resistivities are probably due to the fact that the 12~nm thick Pd film consists of somewhat disconnected islands.
Hydrogen absorption causes the Pd islands to expand which decreases the resistivity between
them.\cite{favier01,dankert02}. Thus, we have a sort of percolation effect and the resistivity is lower in 1~bar H$_2$
than in 40~mbar H$_2$ contrary to bulk PdH$_x$.

\subsubsection{Optical constants of the double layer Pd/Mg}
\label{sssec:optconstdoublelayer}

To investigate the optical properties of the partially interdiffused Pd-Mg top layer, we deposit a layer of
10~nm Mg capped with 10~nm Pd on quartz. This Pd-Mg film is then exposed to hydrogen at 100$^{\circ}$C and
100~bar (together with the thick Mg film covered by Pd (see Fig.~\ref{fig:substrate})). After hydrogenation the
optical properties are determined in the optical gas loading cell at RT. This sample is treated as
consisting of two layers: a Mg-rich Pd-Mg alloy on the substrate covered with a Pd-rich Pd-Mg alloy on top.
As starting values for the fitting procedure we use the PdH$_x$ dielectric function determined above for
the Pd-rich Pd-Mg top layer, and combine it with voids in a Bruggeman effective medium approximation (EMA)
to take surface roughness into account.\cite{aspnes79} A Lorentz-Drude model is used for the second layer,
the Mg-rich Pd-Mg alloy. Ellipsometric data for three angles of incidence (55, 60 and 65$^{\circ}$) and
normal incidence transmisson data are then combined in a multiple data type fit. All data are measured on
the same sample and during fitting the layer optical functions and thicknesses are coupled. The final
iteration results in a fit with a MSE of 8.7. The optical properties of this double layer are then used as
starting values for the top layer of the thicker MgH$_2$ film.

\subsubsection{Dielectric function of MgH$_2$}
\label{sssec:dielectricf}

In the evaluation of the optical properties of MgH$_2$ we analyze ellipsometric and transmission data of a
124~nm thick MgH$_2$ film capped with 12~nm PdH$_x$ (as measured with a mechanical stylus profilometer in
the hydrogenated state). In addition to these data, transmission data of a compositionally identical
film, but with a thickness of 162~nm (when hydrogenated) are included in the modeling. These three data sets
are evaluated in three parallel, coupled models simultaneously. The main features can be modeled using two
Lorentz oscillators at the high energy side of the measured spectra, at 6.4 and 6.9~eV. These oscillators
mark the beginning of the conduction band.

The optical parameters of the capping layer, consisting of the Pd-rich Pd-Mg alloy on top of the Mg-rich
Pd-Mg alloy, are initially fixed to the parameter values obtained in Section~\ref{sssec:optconstdoublelayer}. The only
parameters of the top layers which are allowed to vary are the thicknesses since the diffusion of Pd into
MgH$_2$ may be larger than the 10~nm in the thin Pd/Mg double layer. In the final iteration a global fit is
used in which all LD parameters are allowed to change. The final MSE is 17.17. Table~\ref{table:LDMgH2} gives
the LD parameter values from the final iteration for MgH$_2$, the Pd-rich Pd-Mg cap layer and the Mg-rich
Pd-Mg cap layer.
\begin{table*}

\caption{\label{table:LDMgH2}Lorentz-Drude parameters and their 90\% confidence intervals of a 124~nm
MgH$_2$ film covered with 12~nm Pd obtained from ellipsometric and transmission data (see
Fig.~\ref{fig:psideltaMgH2}). MSE = 17.17 and $\epsilon_{\infty}$ = 1.595 $\pm$ 0.092. All parameters
are in eV.}

\begin{ruledtabular} \begin{tabular}{lllllll}

\multicolumn{7}{l}{MgH$_2$ film, $d=115.9$~nm}\\ \hline

($i$) & $\omega_{p,i}$ & $1/\tau_i$ & ($j$) & $\omega_j$ & $\sqrt{f_j}$ & $\Gamma_j$  \\ \hline
   &&& (1) & 6.4 & 5.516 & 0.6454   \\
   &&& (2) & 6.9 & 7.689 $\pm$ 0.519 & 0.01223 $\pm$ 0.109 \\

\hline \multicolumn{7}{l}{Top cap layer: Pd-rich Pd-Mg alloy (with 53\% voids), $d=15.6$~nm}\\ \hline

($i$) & $\omega_{p,i}$ & $1/\tau_i$ & ($j$) & $\omega_j$ & $\sqrt{f_j}$ & $\Gamma_j$  \\ \hline (1) &
14.35 $\pm$ 0.062 & 1.055 $\pm$ 30.6 & &&& \\
  &&& (1) & 2.964 $\pm$ 0.122 & 5.208 $\pm$ 0.559 & 1.702 $\pm$ 9.03 \\
  &&& (2) & 8.5 & 17.408 $\pm$ 12.2 & 0.01366 $\pm$ 4.96 \\

\hline \multicolumn{7}{l}{Lower cap layer: Mg-rich Pd-Mg alloy, $d=4.1$~nm}\\ \hline

($i$) & $\omega_{p,i}$ & $1/\tau_i$ & ($j$) & $\omega_j$ & $\sqrt{f_j}$ & $\Gamma_j$  \\ \hline (1) &
6.672 $\pm$ 0.865 & 0.003629 $\pm$ 8.29 & &&& \\
  &&& (1) & 3.293 $\pm$ 0.0824 & 24.61 $\pm$ 0.420 & 14.09 $\pm$ 7.06 \\

\end{tabular} \end{ruledtabular} \end{table*}

The plasma frequencies obtained for the two top layers give us a clue about their composition. Since the
plasma frequency of the top layer ($\omega_p = 14.35$~eV) is much larger than the one of the lower cap
layer ($\omega_p = 6.672$~eV), the top layer has a larger charge carrier density and is thus more
metallic than the lower one. This indicates that the top layer is formed by a {\it metallic} Pd-Mg
alloy. The lower layer contains some insulating MgH$_2$ as well. The optical resisitivity (see
Eq.~\ref{eq:optres}) of the top layer is 38~$\mu\Omega$cm compared to an electrical resistivity of
63~$\mu\Omega$cm (at RT) for the total stack as measured after loading. Thus, the top layer is indeed
shunting the resisitivity measurements of MgH$_2$. Since PdH$_x$ has an optical resistivity of
178~$\mu\Omega$cm it is clear that the top layer contains some metallic Mg as well which has a much
lower resistivity (6.5~$\mu\Omega$cm).

The total thickness of the stack obtained from ellipsometry is 130.6~nm after hydrogenation. With the
stylus profilometer we found a thickness of 136~nm. However, as mentioned before the profilometer gives a
value that is too large. Before hydrogenation the thickness was 95~nm. This would mean an increase of
37.5\% instead of the theoretical 32\% volume expansion.

\begin{figure}
\includegraphics[width=8cm]{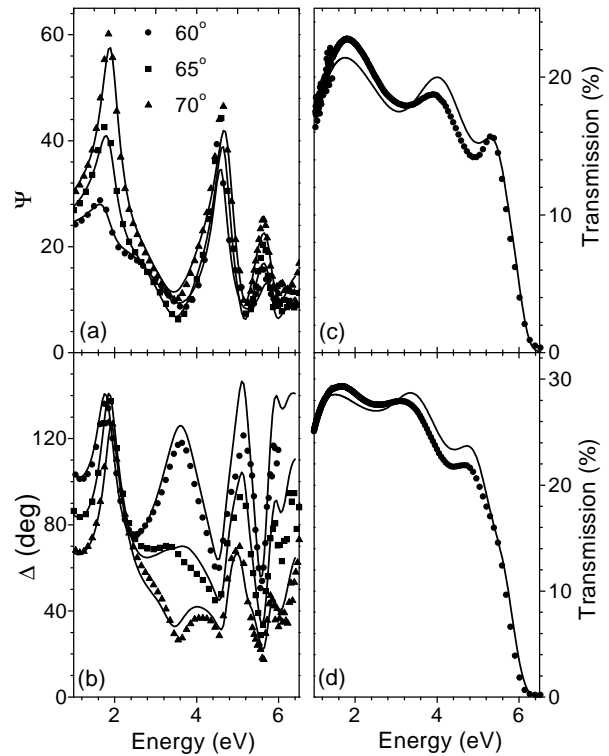}
\caption{Experimental and fitted data for a 124~nm thick MgH$_2$ film covered with 12~nm Pd used to
determine the dielectric function of MgH$_2$ (see Fig.~\ref{fig:eps12MgH2}); (a) ellipsometric data for
$\Psi$, (b) ellipsometric data for $\Delta$, (c) optical transmission (d) transmission of a sample
consisting of 150~nm MgH$_2$/12~nm Pd. All experimental data are modelled simultaneously in the fitting
procedure as described in Sec.~\ref{subsec:ellipsometry}. The mean square error of the fit is 17.17.
\label{fig:psideltaMgH2}}
\end{figure}

\begin{figure}
\includegraphics[width=8cm]{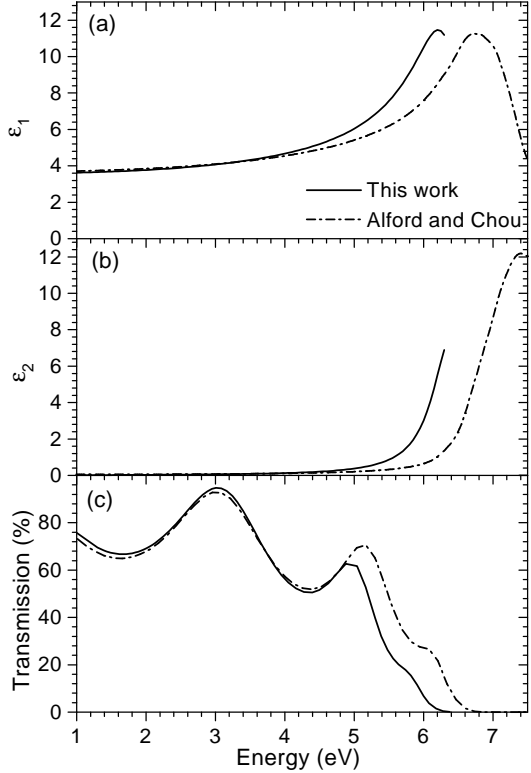}
\caption{Real (a) and imaginary (b) parts of $\epsilon(\omega)$ for MgH$_2$ determined from ellipsometry
and transmission data (see Fig.~\ref{fig:psideltaMgH2}). The real and imaginary parts calculated by
Alford and Chou\cite{alford03} using the GW approximation are shown for comparison. The transmission of
a 100~nm MgH$_2$ film in vacuum calculated with the dielectric functions shown in (a) and (b) is
displayed in the lower panel (c). The theoretical gap is approximately 0.5~eV too large.
\label{fig:eps12MgH2}}
\end{figure}

In Fig.~\ref{fig:psideltaMgH2}(a),(b) the experimental and fitted values of $\Psi$ and $\Delta$ are given,
in (c),(d) the experimental and fitted transmission curves are shown.
Finally, Fig.~\ref{fig:eps12MgH2}(a),(b) shows the real and imaginary part of $\epsilon(\omega)$
obtained for $\alpha$-MgH$_2$.

\section{Discussion} \label{sec:discussion}

In Table~\ref{table:mgh2} both experimental and theoretical values for the band gap of MgH$_2$ are given.
Hartree-Fock (HF) calculations are not included since this method overestimates the
band gap considerably. Very recent and not yet published theoretical work by Herzig\cite{herzig01},
Auluck\cite{auluck02} and Alford and Chou\cite{alford03} on MgH$_2$ is included.
\begin{table}
\caption{\label{table:mgh2}Literature data for the band gap, $E_g$, of MgH$_2$ both experimental (expt.)
and calculated (calc.). In parentheses the smallest direct gap for MgH$_2$ is given.}

\begin{ruledtabular} \begin{tabular}{lllll}
Material & & Method & $E_g$ (eV) & Ref.  \\ \hline

$\alpha$-MgH$_2$ & expt. & UV-absorption & 5.16 & \onlinecite{krasko82} \\
    && XPS & 5.8 & \onlinecite{he90} \\
    && EELS & $>$3.7\footnotemark[1] & \onlinecite{sprunger91}  \\
    && XPS & $>$3.4\footnotemark[1] & \onlinecite{krozer96}  \\
    && Transmission & 6.05\footnotemark[2] & \onlinecite{yamamoto02} \\
    && Ellipsometry/ & 5.6 $\pm$ 0.1 & this work  \\
    && Transmission && \\
    & calc. & LDA & 3.06 & \onlinecite{yu88} \\
    &&  LDA & 3.4 & \onlinecite{pfrommer94} \\
    &&  LDA & 3.45 (4.34) & \onlinecite{herzig01} \\
    &&  LDA & 3.3 & \onlinecite{auluck02} \\
    &&  LDA & 3.10 & \onlinecite{alford03} \\
    &&  GGA  & 3.78 & \onlinecite{haussermann02} \\
    &&  GGA & 4.2 & \onlinecite{vajeeston02}  \\
    && sX-LDA & 5.71 (6.41) & \onlinecite{herzig01} \\
    && GWA & 5.25 (6.11) & \onlinecite{alford03} \\
$\gamma$-MgH$_2$ & calc. & GGA & 4.3 & \onlinecite{vajeeston02}  \\ $\beta$-MgH$_2$ & calc. & APW & 0.23
& \onlinecite{gupta93}  \\
    && GGA & 2.35 & \onlinecite{vajeeston03}  \\

\end{tabular} \end{ruledtabular}

\footnotetext[1]{If there are no charging effects, the band gap would be twice the indicated value.}
\footnotetext[2]{This is the photon energy where the transmission vanishes.}

\end{table}

The band gap of $5.6 \pm 0.1$~eV determined in this work for $\alpha$-MgH$_2$ is close to values
mentioned sporadically in literature. The value 5.16~eV was obtained in an UV-absorption study mentioned
by Krasko.\cite{krasko82} However, details about how this value was obtained have never been published.
The value found by He and Pong\cite{he90} for the average band gap $<E_g>$ is close to ours. This is
rather surprising since $<E_g>$ was obtained in an indirect way from X-ray photoelectron spectroscopy
(XPS) data using Penn's formula.\cite{penn62} Yamamoto {\it et al.}\cite{yamamoto02} measured the
specular optical transmission of thin layers of MgH$_2$ covered by Pd and found that the transmission is
zero at 6.05~eV. However, they did not apply Tauc's method (see Sec.~\ref{subsec:trans}) to the
transmission edge in order to obtain an estimate for the band gap. Furthermore, one should keep in mind
that 6.02~eV is at the detection limit of their Shimadzu spectrophotometer.

As can be seen in Table~\ref{table:mgh2} LDA calculations give a band gap that is systematically too low.
This is a well-known feature of this approximation. Similarly, the GGA used in two other
papers\cite{vajeeston02,haussermann02}
to calculate the density of states of MgH$_2$, underestimates the band gap. Our band gap is
closest to the theoretically calculated gaps of Herzig\cite{herzig01} using screened-exchange-LDA (sX-LDA)
and Alford and Chou\cite{alford03} using the GW approximation (GWA). Furthermore, it is interesting to point out
that our experimentally found gap $E_{\mbox{\tiny{g}}} = 5.6 \pm 0.1$~eV is very close to the difference in
ionization energy between Mg and H: 5.952~eV.

Back in 1955 Ellinger {\it et al.}\cite{ellinger55} determined the refractive index of $\alpha$-MgH$_2$ at
589.3~nm (2.107~eV) and found n = 1.95 and 1.96 for the ordinary and extraordinary rays, respectively. We
find n = 1.94 and k = 7.6$\cdot10^{-3}$ at the same energy which is very close.

Both Auluck\cite{auluck02} using LDA and Alford and Chou\cite{alford03} using LDA and GWA have calculated
the band structure and dielectric function for $\alpha$-MgH$_2$. To obtain the dielectric function only
direct transitions are taken into account. The only difference between the LDA and GWA curves is the energy
position. The dielectric function obtained for $\alpha$-MgH$_2$ with GWA agrees quite well with our
measured values (see Fig.~\ref{fig:eps12MgH2} (a),(b)), while the LDA curve, as expected, is shifted to too low energies.
This indicates that a scissors-operation that shifts the conduction band rigidly with respect to the
valence band works well to correct LDA calculations.\cite{gygi89,delsole93} The energy at which
both $\epsilon_1$ and $\epsilon_2$ exhibit a marked increase is, however, slightly different for the experiment and the GWA
calculation. sX-LDA seems to overestimate the band gap since it gives even larger values for the (in)direct gap than GWA.

Figure~\ref{fig:eps12MgH2}(c) shows the optical transmission for a 100~nm thick MgH$_2$ film in vacuum
as calculated with our experimental dielectric function and the one calculated by Alford and Chou. As
can be seen MgH$_2$ has an intrinsic transparency of about 80\% over the entire visible spectrum. The
difference in energy between the absorption edges of the two transmission spectra and the dielectric
functions is about 0.5~eV. Thus, GWA overestimates the optical gap by 0.5~eV. The direct gap determined
from the band structure using GWA is 6.11~eV. Subtracting 0.5~eV from 6.11~eV gives a value of 5.61~eV
which is within the error margin of our experimentally found gap. More information about the dielectric
function and band structure of MgH$_2$ by Alford and Chou will be published elsewhere.

Yet unpublished calculations indicate that the optical properties of both $\alpha$- and $\gamma$-MgH$_2$ are close to each
other.\cite{alford03,auluck02} Vajeeston {\it et al.}\cite{vajeeston02} find a gap of 4.2 and 4.3~eV for
$\alpha$- and $\gamma$-MgH$_2$, respectively with GGA; Bastide {\it et al.}\cite{bastide80} and Bortz
{\it et al.}\cite{bortz99} have found that the structures of the two different phases are closely related and
that the density is almost the same as well as the H-H distances. Therefore, our dielectric function for $\alpha$-MgH$_2$
is probably a very good approximation for $\gamma$-MgH$_2$ as well. This is important to model the optical properties of
switchable mirrors since in fully hydrogenated Y-Mg alloys Nagengast {\it et al.} found that fcc YH$_3$ coexists with
$\gamma$-MgH$_2$.\cite{nagengast99b} $\beta$-MgH$_2$ seems to be very different both
structurally\cite{bastide80} and optically. Its density is much larger and the calculated band gap for this material
turns out to be considerably smaller than that of both $\alpha$- and $\gamma$-MgH$_2$ (see
Table~\ref{table:mgh2}). The augmented plane wave (APW) calculations underestimate the band gap
considerably, the GGA value of 2.35~eV (again by Vajeeston\cite{vajeeston03}) is much more
reliable. The difference between the calculated band gap using GGA and the one measured for
$\alpha$-MgH$_2$ is 1.4~eV. Assuming that the same scissors-operation can be applied we expect an
experimental gap of 5.7~eV for $\gamma$-MgH$_2$ and 3.75~eV for $\beta$-MgH$_2$.

We now compare MgH$_2$ to related materials such as MgF$_2$, MgO, MgS, MgSe, and other alkaline-earth and
alkali hydrides. In Table~\ref{table:relmgh2} the band gaps of these materials are listed.
\begin{table}
\caption{\label{table:relmgh2}Literature data for the band gap, $E_g$, of materials closely related to
MgH$_2$ both experimental (expt.) and calculated (calc.).}

\begin{ruledtabular} \begin{tabular}{lllll}

Material & & Method & $E_g$ (eV) & Ref.  \\ \hline

MgF$_2$ & expt. & Reflectance & 12.4 & \onlinecite{thomas73}  \\
    & calc. & tight-binding, && \\
    && pseudopotentials & 12.8 & \onlinecite{jouanin76}  \\
MgO & expt. & Reflectance & 7.77 & \onlinecite{roessler67}  \\
    & calc. & LDA & $\sim$5 & \onlinecite{schonberger95}  \\
    && GWA & 7.8 & \onlinecite{shirley98,schonberger95}  \\
MgS & calc. & LDA & 2.6 & \onlinecite{boer98b}  \\
    && LDA (corrected)  & 4.59 & \onlinecite{ching95} \\
MgSe & expt. &  & 5.6 & \onlinecite{strehlow73} \\ MgTe & expt. &  & 4.7, 3.6 & \onlinecite{strehlow73}
\\ CaH$_2$ & expt. & XPS & $\sim$5\footnotemark[1] & \onlinecite{franzen77} \\
    && SRPES & $\sim$5\footnotemark[1] & \onlinecite{weaver84} \\
    & calc. & LDA & 3.32 & \onlinecite{xiao94}  \\
SrH$_2$ & expt. & XPS & $\sim$5\footnotemark[1] & \onlinecite{franzen77} \\ BaH$_2$ & expt. & XPS &
$\sim$5\footnotemark[1] & \onlinecite{franzen77} \\
    && XPS & $>$2.2\footnotemark[2] & \onlinecite{krozer96} \\
LiH & expt. & & 4.99 & \onlinecite{baroni85} \\
    & calc. & GWA & 5.24 & \onlinecite{baroni85} \\
LiD & expt. & & 5.04 & \onlinecite{shirley98} \\
    & calc. & LDA & 2.84 & \onlinecite{shirley98} \\
    & & GWA & 5.37 & \onlinecite{shirley98} \\

\end{tabular} \end{ruledtabular}

\footnotetext[1]{Charging effects are taking into account in the experiments.} \footnotetext[2]{If there
are no charging effects, the band gap would be
    twice as large.}

\end{table}
For all of them it turns out that LDA underestimates the measured gap. Again, as for MgH$_2$, the GW
approximation seems to give a very good agreement between experiment and calculation for MgO, LiH and
LiD. The same is true for the alkali halides.\cite{shirley98} MgH$_2$ and CaH$_2$ seem to be very
similar. Both materials are wide band gap insulators and have a valence band that is predominantly
determined by hydrogen orbitals.\cite{yu88,pfrommer94,xiao94} For MgO and MgS both valence and
conduction bands are determined by the anions (O or S)\cite{boer98b} as well and again the same holds
for the alkali halides.\cite{boer98a}

Since both the valence and conduction band of MgH$_2$ are formed by H states, the band gap of the
alkaline-earth hydrides is expected to be almost independent of the metal as is the case for the alkali
halides.\cite{boer98a,brown70} The remaining dependence results from the influence of the metal-ion on the
lattice parameter, and the influence of hybridization of the conduction band
states with metal-ion $s$- and $d$-states. From XPS experiments by Franzen {\it et al.}\cite{franzen77} it seems that the
onset of transitions in the valence band regions start for CaH$_2$, SrH$_2$ and BaH$_2$ all
at about 2.5~eV. Since charging effects are taking into account, this gives a gap of about 5~eV quite
close to what we have found for $\alpha$-MgH$_2$. It is striking that LiH and LiD also have a gap of 5.0~eV.

\section{Conclusions}
\label{sec:concl}

In this study we use a novel experimental setup for optical transmission and ellipsometry
measurements. This setup facilitates greatly measurements of the optical properties and dielectric function of
metal hydrides in a hydrogen environment. It is possible to control the gas pressure from 1~mbar to 100 bar,
and thus the composition of metal hydrides over a wide range. The temperature, pressure and resistivity are
monitored {\it in situ} during hydrogenation of a sample. We determine the dielectric properties of $\alpha$-MgH$_2$
and PdH$_x$ and find that MgH$_2$ is a transparent, colour neutral insulator with a band gap of $5.6 \pm 0.1$~eV. The
transparency over the whole visible spectrum is $\sim$80\% (for a 100~nm thick film). The experimentally determined
dielectric function in the photon energy range between 1 and 6.5~eV of $\alpha$-MgH$_2$ is in very good agreement
with very recent calculations using the GW approximation. If we assume that calculations of the band gap for $\gamma$- and
$\beta$-MgH$_2$ underestimate the experimental gap by the same amount as in $\alpha$-MgH$_2$ we expect an experimental gap
of 5.7~eV for $\gamma$-MgH$_2$ and 3.75~eV for $\beta$-MgH$_2$.

In a coming publication we shall show that the dielectric function of MgH$_2$ determined here can be
used to explain the large optical absorption (the so-called `black state') of Mg-based alloys where metallic Mg
and insulating MgH$_2$ nanodomains coexist.

\begin{acknowledgments} The authors would like to thank N.J.~Koeman, J.H.~Rector and W.~Lohstroh for their
valuable help with the deposition and characterization of the samples. L.~Jansen is gratefully
acknowledged for constructing the optical gas loading cell and high pressure loading chamber. We are
also very grateful to J.A.~Alford, M.Y.~Chou, P.~Herzig, S.~Auluck and P.~Vajeeston for making available
to us the results of their band structure calculations before publication. This work is part of the
research program of the Stichting voor Fundamenteel Onderzoek der Materie (FOM), financially supported
by the Nederlandse Organisatie voor Wetenschappelijk Onderzoek (NWO).
\end{acknowledgments}

\end{document}